\documentclass[a4]{PoS}

\newcommand{\be}{\begin{equation}}
\newcommand{\ee}{\end{equation}}
\newcommand{\bea}{\begin{eqnarray}}
\newcommand{\eea}{\end{eqnarray}}

\title{Direct photon production at RHIC and LHC energies}

\ShortTitle{Direct photons at RHIC and LHC}

\author{\speaker{O. Linnyk}
                                \\
       Institut f\"ur Theoretische Physik,
        Universit\"at Gie\ss en,
        35392 Gie\ss en,
        Germany\\
        E-mail: \email{linnyk@fias.uni-frankfurt.de}}

\author{E. L. Bratkovskaya\\
        Institut f\"ur Theoretische Physik,
        Universit\"{a}t Frankfurt,
        60438 Frankfurt am Main,
  Germany
}

\author{W. Cassing\\
        Institut f\"ur Theoretische Physik,
        Universit\"at Gie\ss en,
        35392 Gie\ss en,
        Germany
}

\abstract{Direct photon spectra and elliptic flow $v_2$ in heavy-ion
collisions at RHIC and LHC energies are investigated within a
relativistic transport approach incorporating both hadronic and
partonic phases -- the Parton-Hadron-String Dynamics (PHSD). The
results suggest that a large $v_2$ of the direct photons -- as
observed by the PHENIX Collaboration -- signals a significant
contribution of photons produced in interactions of secondary mesons
and baryons in the late stages of the collision. In order to further
differentiate the origin of the direct photon azimuthal asymmetry,
we compare our predictions for the centrality dependence of the
direct photon spectra to the recent measurements by the PHENIX
Collaboration and provide predictions for $Pb+Pb$ collisions at LHC
energies with respect to the direct photon spectra and $v_2(p_T)$
for 0-40\% centrality.}

\FullConference{XXII International Baldin Seminar on High Energy Physics Problems \\
         15-20 September, 2014\\
         JINR, Dubna, Russia}

\begin{document}

\vspace{-0.3cm}
\section{Introduction}
\vspace{-0.3cm}

Real and virtual photons are powerful tools to probe matter under
extreme conditions as created in heavy-ion collisions at
relativistic energies, since electromagnetic radiation is emitted
over the whole collision evolution. The first several fm/c of the
collision are particularly interesting because of the high energy
densities reached (well above the expected phase transition region
to the deconfined phase). The photons interact only
electromagnetically and thus escape to the detector undistorted
through the dense and strongly-interacting initially produced
medium.
Their differential spectra and elliptic flow carry the information
on the properties of the matter produced to the detector.

On the other hand, the measured photons provide a time-integrated
picture of the heavy-ion collision dynamics and are emitted from
every moving charge -- partons or hadrons. Therefore, a multitude of
photon sources has to be differentiated in order to access the
signal of interest. The dominant contributions to the inclusive
photon production are the decays of mesons, mainly pions, eta- and
omega-mesons. The PHENIX and ALICE Collaborations subtract the
``decay photons" from the inclusive photon spectrum using a cocktail
calculation~\cite{PHENIX1,Wilde:2012wc} and obtain the ``direct"
photons.

In particular the direct photons at low transverse momentum
($p_T<3$~GeV) are expected to be dominated by the "thermal" sources,
i.e. the radiation from the strongly interacting Quark-Gluon-Plasma
(sQGP)~\cite{Shuryak:1978ij} as well as secondary meson+meson and
meson+baryon interactions~\cite{Song:1994zs,Li:1998ma}. These
partonic and hadronic channels have been studied within PHSD in
detail in Refs.~\cite{Linnyk:2013hta,Linnyk:2013wma,Linnyk:2013zfa}
at Relativistic-Heavy-Ion-Collider (RHIC) energies. It was found
that the partonic channels constitute up to half of the observed
direct photon spectrum for very central collisions. Other
theoretical calculations find a dominant contribution of the photons
produced in the QGP to the direct photon
spectrum~\cite{Liu:2009kq,Dion:2011pp,Chatterjee:2013naa}.

The low-$p_T$ direct photons probe not only the
temperature~\cite{PHENIX1,Wilde:2012wc,Shen:2013vja} of the produced
QCD-matter, but also its (transport) properties, for instance, the
sheer viscosity. Using the direct photon elliptic flow $v_2$ (a
measure of the azimuthal asymmetry in the photon distribution) as a
viscometer was first suggested by Dusling et al. in
Ref.~\cite{Dusling:2009bc}; this idea was later supported by the
calculations in
Refs.~\cite{Chatterjee:2013naa,Shen:2013vja,Gale:2012rq}. It was
also suggested that the photon spectra and $v_2$ are sensitive to
the collective directed flow of the
system~\cite{vanHees:2014ida,Shen:2014cga} and to the asymmetry
induced by the strong magnetic field (flash) in the very early stage
of the collision~\cite{Bzdak:2012fr,Tuchin:2014pka}.

However, the recent observation by the PHENIX
Collaboration~\cite{PHENIX1} that the elliptic flow $v_2(p_T)$ of
'direct photons' produced in minimal bias Au+Au collisions at
$\sqrt{s_{NN}}=200$~GeV is comparable to that of the produced pions
was a surprise and in contrast to the theoretical expectations and
predictions. Indeed, the photons produced by partonic interactions
in the quark-gluon plasma phase have not been expected to show
considerable flow because they are dominated by the emission in the
initial phase before the elliptic flow fully develops.

In Ref.~\cite{Linnyk:2013hta} we have applied the PHSD approach to
photon production in Au+Au collisions at $\sqrt{s_{NN}}=200$~GeV and
studied the transverse momentum spectrum and the elliptic flow $v_2$
of photons from hadronic and partonic production channels. A
microscopic description of the full collision evolution is done by
the covariant off-shell transport PHSD. The degrees of freedom in
the partonic and hadronic phases are strongly interacting dynamical
quasi-particles and off-shell hadrons, respectively.
Further, it was found in Ref.\cite{Linnyk:2013wma} that the PHSD
calculations reproduce the transverse momentum spectrum of direct
photons as measured by the PHENIX Collaboration in
Refs.~\cite{PHENIXlast,Adare:2008ab}. The centrality dependence of
the thermal photon yield in PHSD was predicted to be $\sim
N_{part}^\alpha$ with the exponent $\alpha=1.5$, which is in a good
agreement with the most recent measurement of
$\alpha=1.48\pm0.08\pm0.04$ by the PHENIX
Collaboration~\cite{Adare:2014fwh}. Furthermore, the PHSD also
described the data on the elliptic flow of inclusive {\em and
direct} photons. The strong $v_2$ of direct photons -- which is
comparable to the hadronic $v_2$ -- in PHSD is attributed to
hadronic channels, i.e. to meson binary reactions which are not
subtracted in the data. As sources for photon production, we
incorporated the interactions of off-shell quarks and gluons in the
strongly interacting quark-gluon plasma (sQGP) ($q+\bar q\to
g+\gamma$ and
 $q(\bar q)+g\to q(\bar q)+\gamma$), the decays of hadrons
($\pi\to\gamma+\gamma$, $\eta\to\gamma+\gamma$,
$\omega\to\pi+\gamma$, $\eta'\to\rho+\gamma$, $\phi\to\eta+\gamma$,
$a_1\to\pi+\gamma$) as well as their interactions
($\pi+\pi\to\rho+\gamma$, $\rho+\pi\to\pi+\gamma$, meson-meson
bremsstrahlung $m+m\to m+m+\gamma$), meson-baryon bremsstrahlung
($m+B\to m+B+\gamma$) and the two-to-two meson+baryon interactions
($\rho+p\to\gamma+p/n$ and $\rho+n\to\gamma+p/n$) .

The photon production via bremsstrahlung in meson-meson and
meson-baryon elastic collisions was found to be an important source
for the direct photon spectra and elliptic flow
simultaneously~\cite{Linnyk:2013hta,Linnyk:2013wma}, where for the
calculation of the photon bremsstrahlung from all elastic
meson-meson and meson-baryon scatterings $m_1 + m_2$, which occur
during the heavy-ion collisions (including $m_i = \pi, \eta, K, \bar
K, K^0, K^*, \bar K^*, K^{*0}, \eta', \omega, \rho, \phi, a_1$), we
have been applying the soft photon approximation. Therefore the
resulting yield of the bremsstrahlung photons depended on the model
assumptions such as (i) the cross section for the meson-meson
elastic scattering (we assumed 10 mb for all meson species), (ii)
incoherence of the individual scatterings and (iii) the soft photon
approximation (i.e. low photon energy and low $\sqrt{s}$ of the
collision). The adequacy of the SPA assumption has been studied in
Ref.~\cite{Eggers:1995jq} and a theoretical uncertainty of up to a
factor of 2 was found.

The results of our calculations so far have been compared to the
data from RHIC. Additionally, here we will provide calculations for
the photon production in $Pb+Pb$ collisions as the energy of
$\sqrt{s_{NN}}=2.76$~TeV. Since the preliminary data of the ALICE
Collaboration~\cite{Wilde:2012wc,Lohner:2012ct} indicate a
significant direct photon signal at low $p_T$ with a large elliptic
flow at LHC energies, a differential comparison of our calculations
with the final data will be mandatory.

\vspace{-0.3cm}
\section{PHSD}
\vspace{-0.3cm}

To address the photon production in a hot and dense medium -- as
created in heavy-ion collisions -- we employ an up-to-date
relativistic transport model, i.e. the Parton Hadron String
Dynamics~\cite{CasBrat} (PHSD) that incorporates the explicit
partonic phase in the early reaction phase. Within PHSD, one solves
generalized transport equations on the basis of the off-shell
Kadanoff-Baym equations for Greens functions in phase-space
representation (in first order gradient expansion, beyond the
quasiparticle approximation).  The approach consistently describes
the full evolution of a relativistic heavy-ion collision from the
initial hard scatterings and string formation through the dynamical
deconfinement phase transition to the quark-gluon plasma (QGP) as
well as hadronization and to the subsequent interactions in the
hadronic phase.
In the hadronic sector PHSD is equivalent to the
Hadron-String-Dynamics (HSD) transport approach \cite{Brat97} that
has been used for the description of $pA$ and $AA$ collisions from
SIS to RHIC energies and has lead to a fair reproduction of hadron
abundances, rapidity distributions and transverse momentum spectra.
The description of quarks and gluons in PHSD is based on a dynamical
quasiparticle model for partons matched to reproduce lattice QCD
results in thermodynamic equilibrium (DQPM). The DQPM describes QCD
properties in terms of single-particle Green's functions (in the
sense of a two-particle irreducible approach) and leads to the
notion of the constituents of the sQGP being effective
quasiparticles, which are massive and have broad spectral functions
(due to large interaction rates). The transition from partonic to
hadronic degrees of freedom in PHSD is described by covariant
transition rates for the fusion of quark-antiquark pairs to mesonic
resonances or three quarks (antiquarks) to baryonic states.
The PHSD transport approach provides a good description of the data
on bulk properties~\cite{Konchakovski:2014fya} as well as hard
probes~\cite{Linnyk:2011vx} for a wide range of energies up to the
LHC.

\vspace{-0.3cm}
\section{Direct photon production}
\vspace{-0.3cm}

We consider the following sources of direct photons:

1) Photons radiated by quarks in the interaction with other quarks
and gluons in the two-to-two reactions:
\begin{eqnarray}
q + \bar{q} \rightarrow g + \gamma , \nonumber\\
q/\bar{q} + g \rightarrow q/\bar{q} + \gamma .  \nonumber
\label{quark22a}
\end{eqnarray}
The implementation of the photon production by the quark and gluon
interactions in the PHSD is based on the off-shell cross sections
for the interaction of massive dynamical quasi-particles as
described in~\cite{Linnyk:2013hta,olena2010}.  In addition, photon
production in the bremsstrahlung reactions $q+q/g\to q +q/g +
\gamma$ should be incorporated.

2) All colliding hadronic charges (meson, baryons) can also radiate
photons by the bremsstrahlung process:
\begin{eqnarray}
m+m\to m +m + \gamma, \label{mmBr} \\
m+B\to m +B + \gamma. \label{mBbr}
\end{eqnarray}
The processes (\ref{mmBr}) have been calculated within the PHSD in
Refs.~\cite{Linnyk:2013hta,Bratkovskaya:2008iq}, while the $m+B$
bremsstrahlung (\ref{mBbr}) reactions have been added in
Ref.~\cite{Linnyk:2013wma}. The implementation of photon
bremsstrahlung from hadronic reactions in transport approaches so
far has been based on the 'soft photon' approximation
(SPA)~\cite{Gale87}, which relies on the assumption that the
radiation from internal lines is negligible and the strong
interaction vertex is on-shell; this is valid only at low energy
(and $p_T$) of the produced photon.

3) Additionally, the photons can be produced in specific binary
hadronic collisions. We consider the direct photon production in the
following $2\to2$ meson+meson collisions
\begin{eqnarray}
\pi + \pi \rightarrow \rho + \gamma , \nonumber\\
\pi + \rho \rightarrow \pi + \gamma ,  \nonumber \label{22}
\end{eqnarray}
accounting for all possible charge combinations. The implementation
of these reactions has been described in
Refs.~\cite{Linnyk:2013hta,Bratkovskaya:2008iq}.

\vspace{-0.3cm}
\section{Results}
\label{sect:results} \vspace{-0.3cm}

\begin{figure*}
  \begin{minipage}[b]{0.495\textwidth}
    \centering
\includegraphics[width=\textwidth]{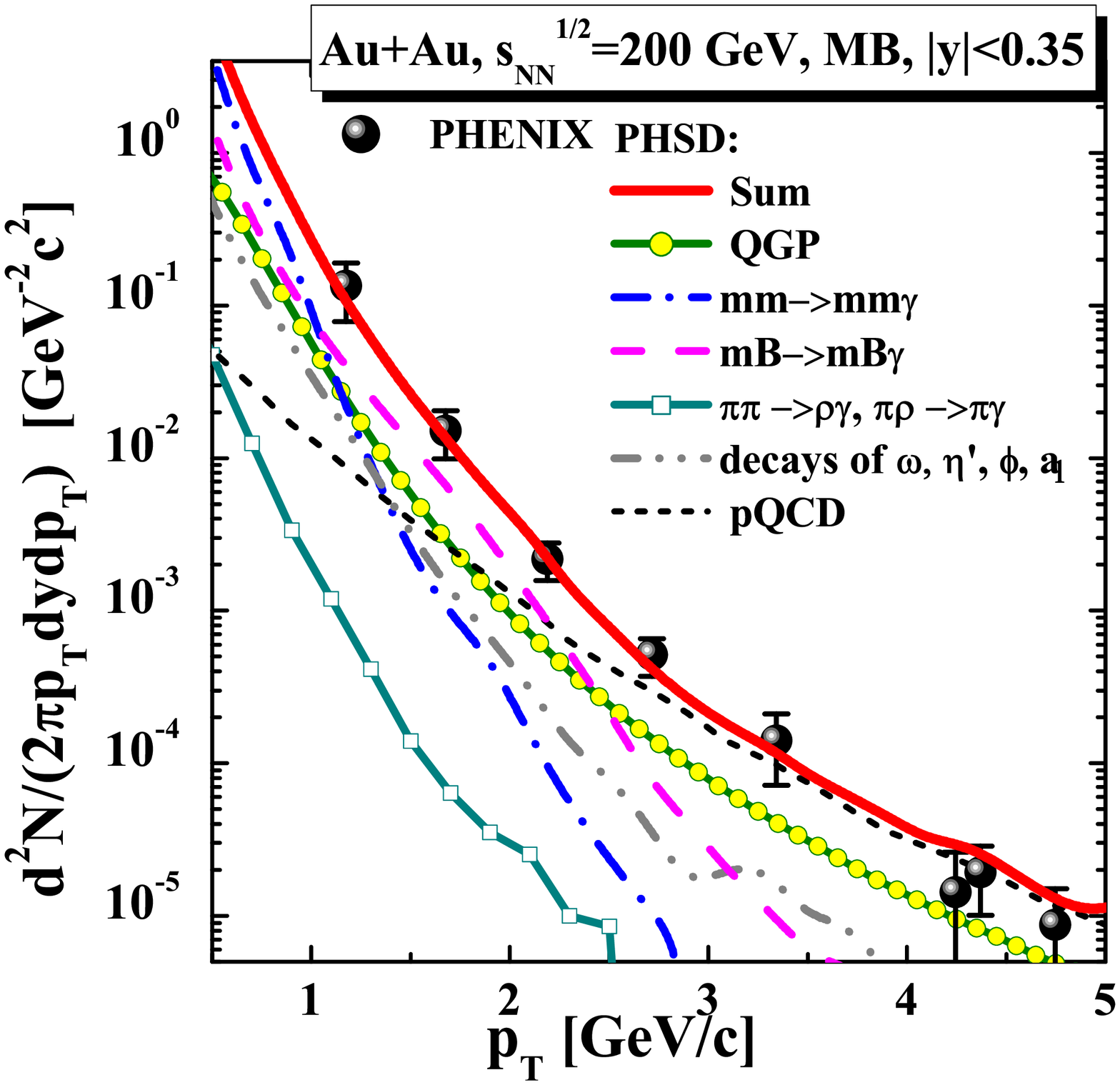}
\caption{PHSD results for the spectrum of direct photons produced in
0-40\% most central Au+Au collisions at $\sqrt{s_{NN}}=200$~GeV as a
function of the transverse momentum $p_T$ at mid-rapidity $|y|<
0.5$. The data of the PHENIX Collaboration are taken from
Ref.~\protect{\cite{Adare:2008ab}}.} \label{spectrarhic}
  \end{minipage}
  \hspace{0.02\textwidth}
  \begin{minipage}[b]{0.495\textwidth}
    \centering
\includegraphics[width=\textwidth]{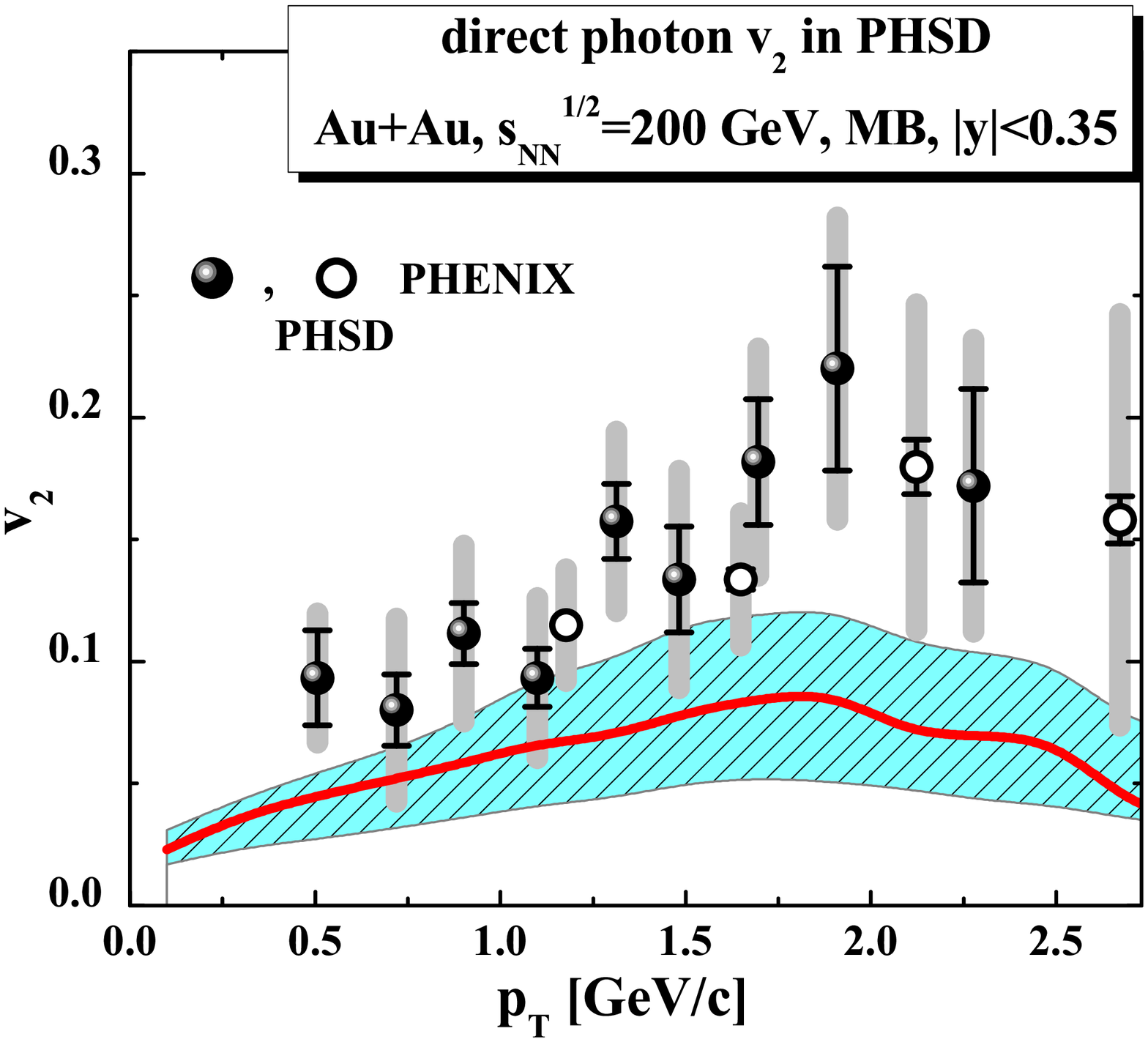}
\caption{  Elliptic flow $v_2$ versus transverse momentum $p_T$ for
the direct photons produced in the minimal bias Au+Au collisions at
$\sqrt{s_{NN}}=200$~GeV calculated within the PHSD (solid red line);
the blue band reflects the uncertainty in the modeling of the cross
sections for the individual channels. The data of the PHENIX
Collaboration are from Ref.~\protect{\cite{PHENIX1}}.}
\label{rhicV2}
  \end{minipage}
\end{figure*}

The results for the direct photon spectrum as a sum of partonic as
well as hadronic sources for the photons produced in 0-40\% central
Au+Au collisions at $\sqrt{s_{NN}}=200$~GeV is presented in
Fig.~\ref{spectrarhic} as a function of the transverse momentum
$p_T$ at mid-rapidity $|y|< 0.5$. The calculated channel
decomposition of the spectrum is presented in Fig.~\ref{spectrarhic}
by the lines of various styles.  The measured transverse momentum
spectrum $dN/dp_T$ (given by the filled circles) is reproduced well
by the sum of partonic and hadronic sources (red solid line).

We find that the radiation from the sQGP constitutes slightly less
than half of the observed number of photons. The radiation from
hadrons and their interaction -- which are not measured separately
so far -- give a considerable contribution at low transverse
momentum. The dominant hadronic sources are the meson decays and the
meson-meson bremsstrahlung. While the former (e.g. the decays of
$\omega$, $\eta$', $\phi$ and $a_1$ mesons) can be subtracted from
the spectra once the mesonic yields are determined independently by
experiment, the reactions $\pi+\rho\to\pi+\gamma$, $\pi+\pi\to
\rho+\gamma$, $\rho+p/n\to n / p+\gamma$ and the meson-meson and
meson-baryon bremsstrahlung can be separated from the partonic
sources only using theoretical models.

The azimuthal momentum distribution of the emitted particles is
commonly expressed in the form of Fourier series as
\be
E\frac{d^3N}{d^3p}= \nonumber \\
\frac{d^2N}{2\pi p_Tdp_Tdy}\left(1+\sum^\infty_{n=1} 2v_n(p_T) \cos
[n(\psi-\Psi_n)]\right),\ \ \ \label{eqvn} \ee
where $v_n$ is the magnitude of the $n$th order harmonic term
relative to the angle of the initial-state spatial plane of symmetry
$\Psi_n$. One should take into account event-by-event fluctuations
with respect to the event plane $\Psi_{EP}$. We calculate the $v_3$
coefficients with respect to $\Psi_3$ as $v_3\{\Psi_3\} = \langle
\cos(3[\psi-\Psi_3])\rangle/\rm{Res}(\Psi_3)$. The event plane angle
$\Psi_3$ and its resolution $\rm{Res}(\Psi_3)$ are calculated as
described in Ref.~\cite{{Adare:2011tg}}.
We recall that the second flow coefficient $v_2$ carries information
on the interaction strength -- and thus on the state of matter and
its properties -- at the space-time point, from which the measured
particles are emitted.
%

About a decade ago, the WA98 Collaboration has measured the elliptic
flow $v_2$ of photons produced in $Pb+Pb$ collisions at the beam
energy of $E_{beam}=158$~AGeV~\cite{Aggarwal:2004zh}, and it was
found that the $v_2(\gamma^{incl})$ of the low-transverse-momentum
inclusive photons was equal to the $v_2(\gamma^{\pi})$ of pions
within the experimental uncertainties. This observation lead to the
conclusion that either (Scenario a:) the contribution of the direct
photons to the inclusive ones is negligible in comparison to the
decay photons, mainly the $\pi^0$ decay products, or (Scenario 2:)
the elliptic flow of the direct photons is comparable in magnitude
to the $v_2(\gamma^{incl})$,  $v_2(\gamma^{decay})$ and
$v_2({\pi})$.
\begin{figure*} 
\hspace{1.5cm}
\includegraphics[width=0.8\textwidth]{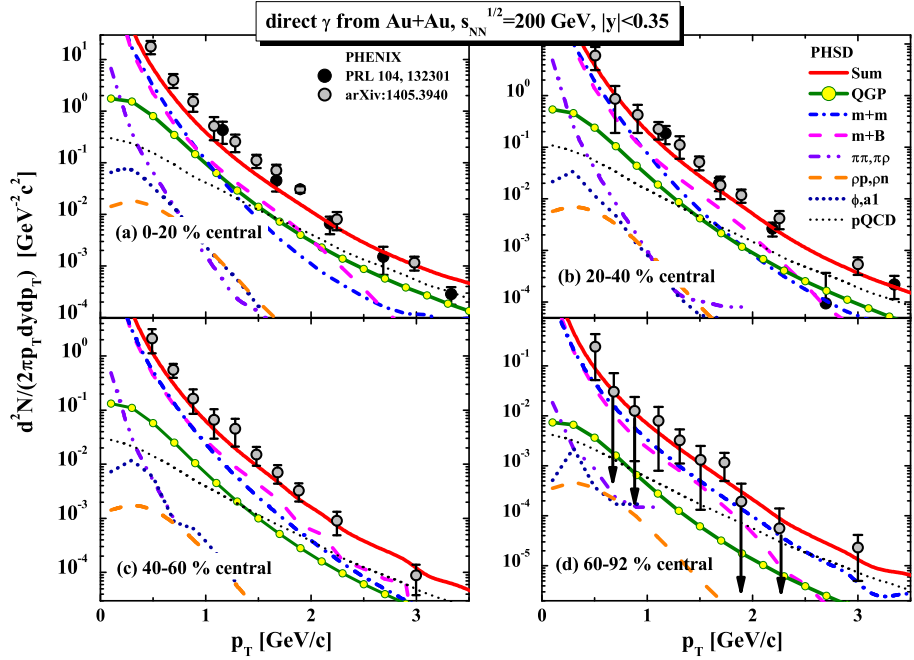}
\caption{  Contribution of the photon production in the two-to-two
$\rho$+nucleon interaction (orange dash lines) to the total direct
photon spectra (red lines) at the top RHIC energy of
$\sqrt{s_{NN}}=200$~GeV at different centralities. The dominant
sources are the photons from the QGP and from the hadronic
two-to-three bremsstrahlung processes. The theory lines -- except
the $\rho$+nucleon channel -- are taken from the PHSD predictions
published in \protect{\cite{Linnyk:2013wma}}. The PHENIX data are
from Ref.~\protect{\cite{Adare:2008ab,Adare:2014fwh}}. \label{222} }
\end{figure*}

However, in view of the WA98 measurement of the direct photon
spectrum, which we described above, there is a significant finite
yield of direct photons at low transverse momentum. Thus the
scenario 1 can be ruled out. Furthermore, the observed direct
photons of low $p_T$ must have a significant elliptic anisotropy
$v_2$ of the same order of magnitude as the hadronic flow. This
scenario points towards hadronic sources of the direct photons.
Thus, the interpretation~\cite{Bratkovskaya:2008iq,Liu:2007zzw} of
the low-$p_T$ direct photon yield measured by WA98 -- as dominantly
produced by the bremsstrahlung process in the mesonic collisions
$\pi+\pi\to\pi+\pi+\gamma$ -- is in accord also with the data on the
photon elliptic flow $v_2(\gamma^{incl})$.
\begin{figure} \centering
\includegraphics[width=0.5\textwidth]{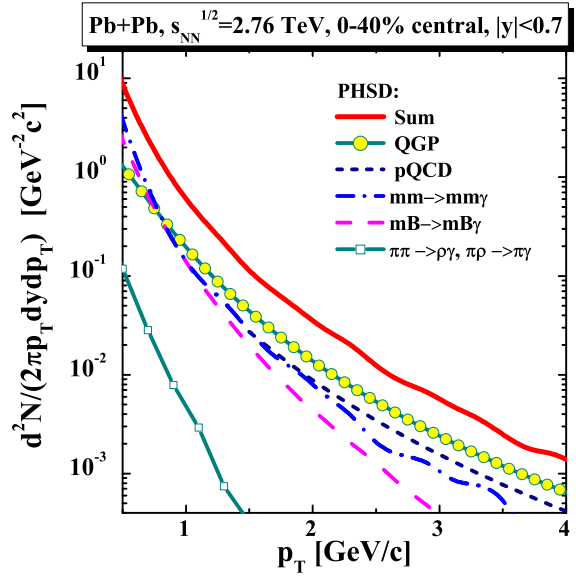}
\caption{  Yield of direct photons in Pb+Pb collisions at the
invariant energy $\sqrt{s_{NN}}=2.76$~TeV for 0-40\% centrality as
predicted within the PHSD. \label{spectralhc} }
\end{figure}

Let us note that the same conclusions apply also to the most recent
studies of the photon elliptic flow at RHIC and LHC. The PHENIX and
ALICE Collaborations have measured the inclusive photon $v_2$ and
found that at low transverse momenta it is comparable to the
$v_2(p_T)$ of decay photons as calculated in cocktail simulations
based on the known mesonic $v_2(p_T)$. Therefore, (a) either the
yield of the direct photons to the inclusive ones is not
statistically significant in comparison to the decay photons or (b)
the elliptic flow of the direct photons must be as large as
$v_2(\gamma^{decay})$ and $v_2(\gamma^{incl})$.

In the PHSD, we calculate the direct photon $v_2(\gamma^{dir})$ by
building the weighted sum of the channels, which are not subtracted
by the data-driven methods, as follows: the photons from the
quark-gluon plasma, from the initial hard parton collisions (pQCD
photons), from the decays of short-living resonances ($a$-meson,
$\phi$-meson, $\Delta$-baryon), from the $2\to2$ channels
($\pi+\rho\to\pi+\gamma$, $\pi+\pi\to\rho+\gamma$), and from the
bremsstrahlung in the elastic meson+meson and meson+baryon
collisions ($m+m\to m+m+\gamma$, $m+B\to m+B+\gamma$). We calculate
the direct photon $v_2$ (in PHSD) by summing up the elliptic flow of
the individual channels contributing to the direct photons, using
their contributions to the spectrum as the relative $p_T$-dependent
weights, $w_i(p_T)$, i.e.
\be \label{dir2}  v_2 (\gamma^{dir}) = \sum _i  v_2 (\gamma^{i}) w_i
(p_T) =  \frac{\sum _i  v_2 (\gamma^{i}) N_i (p_T)}{\sum_i N_i
(p_T)}. \ee
The results for the elliptic flow $v_2(p_T)$ of direct photons
produced in $Au+Au$ collisions at the top RHIC energy are shown in
Fig.~\ref{rhicV2}. According to our calculations of the direct
photon spectra, almost a half of the direct photons measured by
PHENIX stems from the collisions of quarks and gluons in the
deconfined medium created in the initial phase of the collision. The
photons produced in the QGP carry a very small $v_2$ and lead to an
overall direct photon $v_2$ about a factor of 2 below the pion
$v_2(\pi)$ even though the other channels in the sum (\ref{dir2})
have large elliptic flow coefficients $v_2$ of the order of
$v_2(\pi)$ (cf. Ref.~\cite{Linnyk:2013hta}).

Indeed, the parton collisions -- producing photons in the QGP --
take place throughout the evolution of the collision but the
collision rate falls rapidly with time and thus the production of
photons from the QGP is dominated by the early times (cf. Fig.~7 in
Ref.~\cite{Linnyk:2013hta}). As a consequence, the elliptic flow
`picked up' by the photons from the parent parton collisions
saturates after about 5~fm/c and reaches a relatively low value of
about $0.02$, only. We note that a delayed production of charges
from the strong gluon fields (`glasma'~\cite{McLerran:2011zz}) might
shift the QGP photon production to somewhat later times when the
elliptic flow is built up more. However, we cannot quantitatively
answer whether the additional evolution in the pre-plasma state
could generate considerable additional $v_2$.
\begin{figure}
\includegraphics[width=0.49\textwidth]{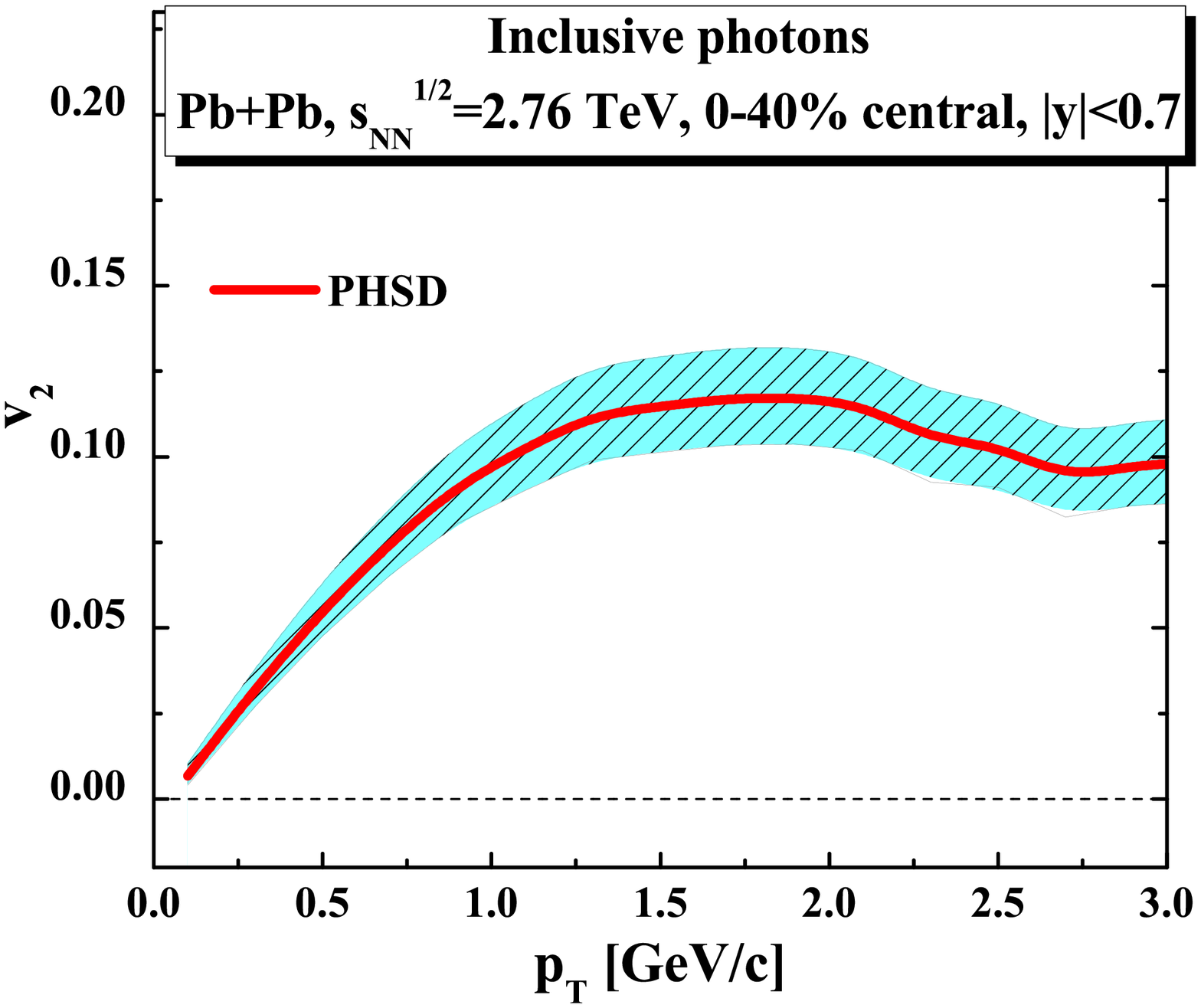}
\includegraphics[width=0.5\textwidth]{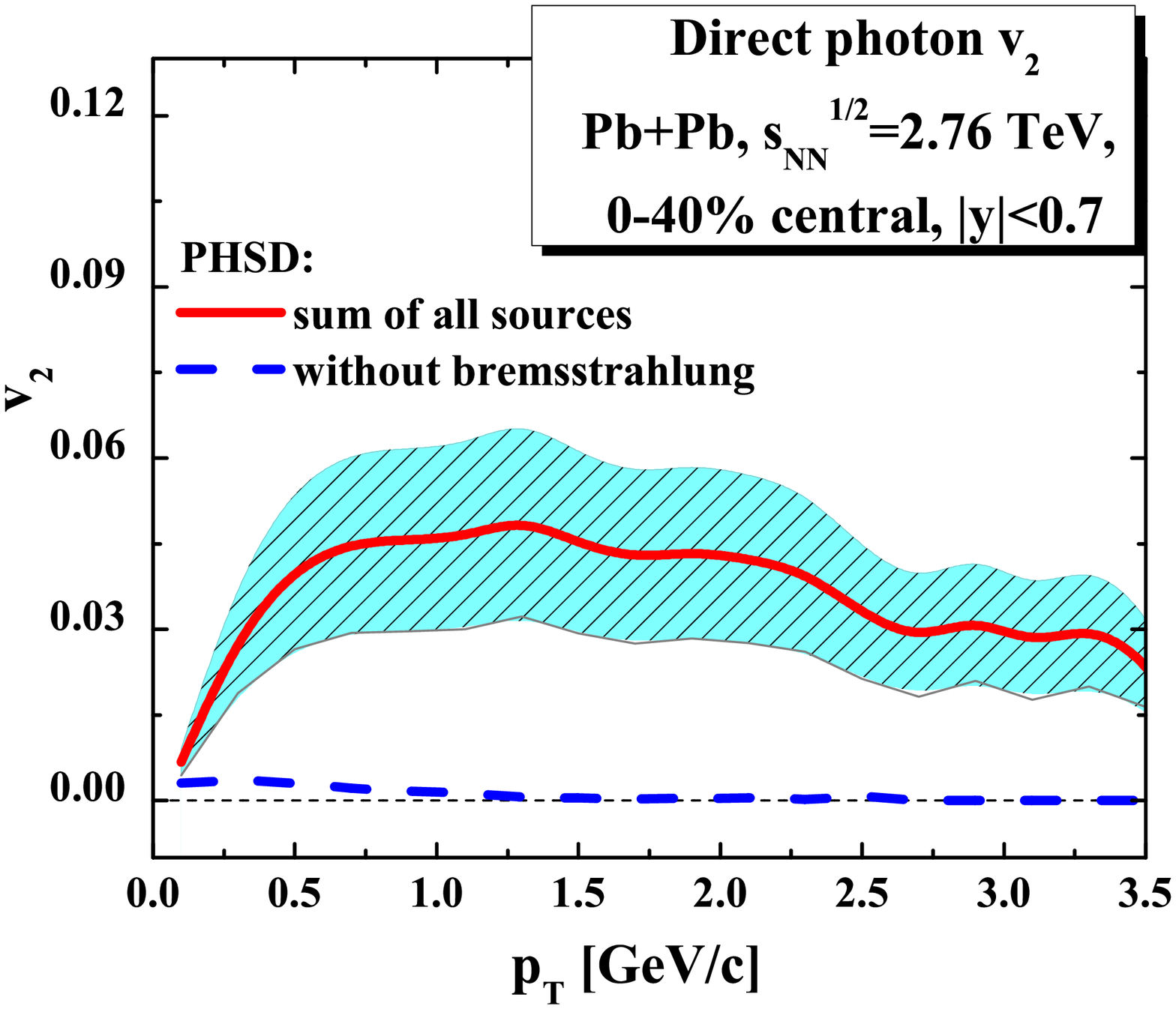}
\caption{ {\bf Left:} Elliptic flow $v_2$ versus transverse momentum
$p_T$ for the inclusive photons produced in 0-40\% central Pb+Pb
collisions at $\sqrt{s_{NN}}=2.76$~TeV as predicted by the PHSD
(solid red line); the blue error band reflects the finite statistics
and the uncertainty in the modeling of the cross sections for the
individual channels. {\bf Right:} Elliptic flow $v_2$ versus
transverse momentum $p_T$ for the direct photons produced in 0-40\%
central Pb+Pb collisions at $\sqrt{s_{NN}}=2.76$~TeV as predicted by
the PHSD (solid red line); the blue error band is dominated by the
uncertainty in the modeling of the cross sections for the individual
channels.} \label{inclv2lhc} \label{lhcv2dir}
\end{figure}

In addition, the centrality dependence of the direct photon spectra
and flow has been investigated. The recent measurements by the
PHENIX Collaboration~\cite{Adare:2014fwh} confirm the predictions
within the PHSD from Ref.~\cite{Linnyk:2013wma}. We present our
predictions and the data in Fig.~\ref{222}. The centrality
dependence of the integrated thermal photon yield in PHSD was found
to scale as $N_{part}^\alpha$ with the exponent $\alpha=1.5$, which
is in a good agreement with the most recent measurement of
$\alpha=1.48\pm0.08\pm0.04$ by the PHENIX
Collaboration~\cite{Adare:2014fwh}.

Centrality dependence of the direct photon elliptic flow $v_2$ was
also calculated within the PHSD in Ref.~\cite{Linnyk:2013wma} and
confirmed (within the error bars) by the PHENIX Collaboration
measurements in Ref.~\cite{Adare:2008ab,Adare:2014fwh}. Thus the
observed centrality dependence of the elliptic flow is in agreement
with the interpretation of the direct photons having a hadronic
origin (in particular from the bremsstrahlung in meson+meson and
meson+baryon collisions), which is stronger in more peripheral
collisions.


In Fig.~\ref{spectralhc} we show the calculated direct photon yield
in Pb+Pb collisions at the invariant energy $\sqrt{s_{NN}}=2.76$~TeV
for 0-40\% centrality. Comparing the theoretical predictions to the
preliminary data of the ALICE Collaboration from
Ref.~\cite{Wilde:2012wc}, we find an overall agreement with the data
within about a factor of 2 in the range of transverse momenta $p_T$
from 1 to 4 GeV. On the other hand, the calculations tend to
underestimate the preliminary data in the low-$p_T$ region and to
underestimate slightly the highest-$p_T$ points. One cannot exclude
the existence of another yet unknown source of direct photons at low
$p_T$. However, the significance of the comparison is not robust
untill the final data will be available.

We finally present our predictions for the elliptic flow of
inclusive and direct photons produced in $Pb+Pb$ collisions at the
energy of $\sqrt{s_{NN}}=2.76$~TeV at the LHC within the acceptance
of the ALICE detector. Fig.~\ref{inclv2lhc} (left hand side)
presents predictions for the elliptic flow $v_2$ versus transverse
momentum $p_T$ for the {\em inclusive} photons produced in 0-40\%
central Pb+Pb collisions at $\sqrt{s_{NN}}=2.76$~TeV (solid red
line) with the blue error band reflecting the finite statistics and
the theoretical uncertainty in the modeling of the cross sections
for the individual channels. Finally, the elliptic flow $v_2(p_T)$
of {\em direct} photons produced in 0-40\% central Pb+Pb collisions
at $\sqrt{s_{NN}}=2.76$~TeV  --as predicted by the PHSD (solid red
line) -- is shown in Fig.~\ref{lhcv2dir} (right hand side); the blue
error band is dominated by the uncertainty in the modeling of the
cross sections for the individual channels. The lines presented in
Fig.~\ref{inclv2lhc} will have to be compared to the future
experimental data in order to shed further light on the direct
photon sources and production mechanisms.

\vspace{-0.3cm}
\section{Summary and outlook}
\vspace{-0.3cm}

In this contribution we have calculated the momentum spectra and the
elliptic flow $v_2$ of direct photons produced in Au+Au collisions
at $\sqrt{s_{NN}}=200$~GeV and in $Pb+Pb$ collisions at
$\sqrt{s_{NN}}=2.76$~TeV using the microscopic PHSD transport
approach. For photon production we have incorporated the
interactions of quarks and gluons in the strongly interacting
quark-gluon plasma (sQGP) ($q+\bar q\to g+\gamma$ and
 $q(\bar q)+g\to q(\bar q)+\gamma$), the photon production in the hadronic decays
($\pi\to\gamma+\gamma$, $\eta\to\gamma+\gamma$,
$\omega\to\pi+\gamma$, $\eta'\to\rho+\gamma$, $\phi\to\eta+\gamma$,
$a_1\to\pi+\gamma$) as well as the interactions
($\pi+\pi\to\rho+\gamma$, $\rho+\pi\to\pi+\gamma$, and the
bremsstrahlung radiation $m+m/B\to m+m/B+\gamma$) of mesons and
baryons produced throughout the evolution of the collision.

We find that the PHSD calculations reproduce the transverse momentum
spectrum of direct photons as measured by the PHENIX Collaboration
in Refs.~\cite{PHENIXlast,Adare:2008ab}. The calculations reveal the
channel decomposition of the observed direct photon spectrum and
show that the photons produced in the QGP constitute at most about
50\% of the direct photons with the rest being distributed among the
other channels: mesonic interactions, decays of massive hadronic
resonances and the initial hard scatterings.
Our calculations demonstrate that the photon production in the QGP
is dominated by the early phase (similar to hydrodynamic models) and
is localized in the center of the fireball, where the collective
flow is still rather low, i.e. on the 2-3 \% level, only.
Thus, the strong $v_2$ of direct photons - which is comparable to
the hadronic $v_2$ - in PHSD is attributed to hadronic channels,
i.e. to meson and baryon binary reactions. On the other hand, the
strong $v_2$ of the 'parent' hadrons, in turn, stems from the
interactions in the QGP via collisions and the partonic mean-filed
potentials. Accordingly, the presence of the QGP shows up
'indirectly' in the direct photon elliptic flow.
In the future, we plan to (i) go beyond the soft photon
approximation (SPA) in the calculation of the bremsstrahlung
processes $meson+meson\to meson+meson+\gamma$, $meson+baryon\to
meson+baryon+\gamma$ and (ii) will quantify the suppression at low
$p_T$ due the Landau-Migdal-Pomeranchuk (LMP) effect.

The centrality dependence of the direct photon production the
potential to further clarify the direct photon production
mechanisms. We find a good agreement between the PHENIX measurements
and PHSD calculations. In particular, the integrated thermal photon
yield in PHSD was predicted to scale as $N_{part}^\alpha$ with the
exponent $\alpha=1.5$, which is in good agreement with the most
recent measurement of $\alpha=1.48\pm0.08\pm0.04$ by the PHENIX
Collaboration~\cite{Adare:2014fwh}. This observation supports the
conclusion that the low transverse momentum direct photons have a
strong contribution from the binary hadronic photon production
sources, such as the $meson+meson$ and $meson+baryon$
bremsstrahlung. It will be important to investigate experimentally
the scaling of the direct photon yield and flow with the number of
participating nucleons $N_{part}$ at LHC, too.


\vspace{-0.3cm}
\section*{Acknowledgments}
\vspace{-0.3cm}

 We gratefully acknowledge fruitful discussions
with C. Gale, C.~M.~Ko, L. Mc Lerran, K. Eskola , I. Helenius,
I.~Tserruya, N.~Xu, C. Klein-Boesing, R. Rapp, H. van Hess, J.
Stachel, U. Heinz, I. Selyuzhenkov, G. David, M. Thoma, K. Reygers,
C. Shen, A. Drees, Gy. Wolf, B. Bannier and F. Bock.


\end{document}